\documentclass[reprint,superscriptaddress,amsmath,amssymb,aps]{revtex4-2}

\usepackage{graphicx}
\usepackage{dcolumn}
\usepackage{bm}

\usepackage{ulem}

\usepackage{bm}

\usepackage{algorithm}
\usepackage[noend]{algpseudocode} 
\usepackage{amssymb}

\usepackage{tabularx}
\usepackage{graphicx}
\usepackage{dcolumn}
\usepackage{bm}
\usepackage{amsmath}
\usepackage{xcolor} 
\usepackage[pagebackref]{hyperref} 
\hypersetup{
	colorlinks   = true, 
	urlcolor     = blue, 
	linkcolor    = blue, 
	citecolor   = blue 
}
\usepackage{graphicx}
\usepackage{physics}
\usepackage{mathtools}
\usepackage{dsfont}
\usepackage{natbib}

\begin{document}
\title{Reconfigurable circuit for mode tunable topological quantum structured light
}

\author{Pedro Ornelas}
\address{School of Physics, University of the Witwatersrand, Private Bag 3, Wits 2050, South Africa}

\author{Tatjana Kleine}
\address{School of Physics, University of the Witwatersrand, Private Bag 3, Wits 2050, South Africa}

\author{Andr\'e G. de Oliveira}
\address{Department of Physics and SUPA, University of Strathclyde, Glasgow G4 0NG, UK}

\author{Carmelo Rosales-Guzmán}
\address{Centro de Investigaciones en Optica A.C., Loma del Bosque 115, Colonia Lomas del Campestre, L\'eon, 37150, Guanajuato, Mexico.}

\author{Andrew Forbes}
\address{School of Physics, University of the Witwatersrand, Private Bag 3, Wits 2050, South Africa}

\author{Isaac Nape}
\email{isaac.nape@wits.ac.za}
\address{School of Physics, University of the Witwatersrand, Private Bag 3, Wits 2050, South Africa}

\vspace{10pt}

\begin{abstract}
\noindent Structured light in the quantum regime has garnered considerable attention due to the opportunities it offers when mixing light’s internal degrees of freedom, for high-dimensional and multi-dimensional quantum states of light. A popular example is to harness polarisation and spatial entangled photons with a shared topological invariant that is robust against numerous families of noisy quantum channels. Yet, producing such states with high purity and adaptability remains challenging. Here we introduce a compact, self-locking Mach–Zehnder interferometer that integrates digital spatial light modulators with static beam displacers to map spatial-mode entanglement from a parametric down-conversion source onto topological entanglement with high fidelity. The device also mimics the action of a reprogrammable controlled-unitary gate, digitally driven by the spatial light modulator. This approach is an enabling platform and provides a practical route to generating reliable, high-purity quantum-structured light with topological features, both at the single-photon level and as entangled states, a direction of growing topical interest.
\end{abstract}

%
%
%
%
\vspace{2pc}
\maketitle
 
\section*{Introduction}
Structured light, engineered across multiple degrees of freedom, has gained prominence and is enabling the generation of exotic states in both the classical and quantum regimes \cite{he2022towards, forbes2025progress}. A notable trend is the mixing of spatial and polarisation degrees of freedom to enhance the performance of quantum protocols, with applications in communications \cite{barreiro2008beating}, metrology \cite{d2013photonic,barboza2022ultra}, and computing \cite{wang2024polarization}, to name a few. On the other hand, structuring these degrees of freedom has opened up new research avenues such as the possibility of probing complex quantum coalescence phenomena with nonclassical features \cite{schiano2024engineering, hong2023hong}, and, more recently, for creating particle-like optical topologies with single- photons \cite{ma2025nanophotonic},  quantum entangled photons \cite{ornelas2024non} and both \cite{koni2025dual}. Remarkably, topological features of photons in the quantum regime offer a potential route towards noise-resilience and robust quantum information processing \cite{ornelas2025topological, de2025quantum, wang2025topological, guo2026topological}. 

Generating quantum topological states often relies on methods that create hybrid entanglement between spatial and spin degrees of freedom with spin-orbit coupling optics \cite{nagali2010generation, karimi2010spin, forbes2019quantum, giovannini2011resilience} or digital modulation with SLMs \cite{nape2022all, xu2025hybrid}. 
Moreover, the ability to tune between different orbital angular momentum states (as illustrated conceptually in Fig.~\ref{fig:schematic}(a), enables the possibility of switching between a wide range of independent topological states \cite{liu2020multidimensional, zhao2024integrated} which can potentially serve as a large encoding alphabet for quantum communications and computing protocols. However, generating arbitrary two-photon topological states with high fidelity and which demonstrate genuine nonlocality, i.e., convincing violations of the Bell inequality or equivalent, remains challenging. 


In this work, we introduce a compact scheme for generating hybrid entangled states that can be encoded with arbitrary topological skyrmionic orders as shown in Fig.~\ref{fig:schematic} (a). We show high Bell-inequality violations across a wide range of topological states, indicating the generation of tuneable, high-purity and genuinely non-local states. Our approach is modular and operationally shown to encode an adaptable controlled unitary rotation gate in the spatial-polarization basis. Integrated into a spatial-mode entanglement setup, the self-locking interferometer that achieves this, is used to map photon pairs produced via spontaneous parametric down-conversion (SPDC) onto target topological entangled states, which we demonstrate by generating states characterized with topological numbers ranging from $N=-5$ to $N=5$ (11 topological classes). All the states are produced with fidelities greater than 80\%. Our approach provides a practical pathway to reliable, high-purity, structured photons with adaptable topological orders.
\vspace{0.25in}
\section*{Concept and implementation}
\begin{figure*}[t]
	\centering
\includegraphics[width=\linewidth]{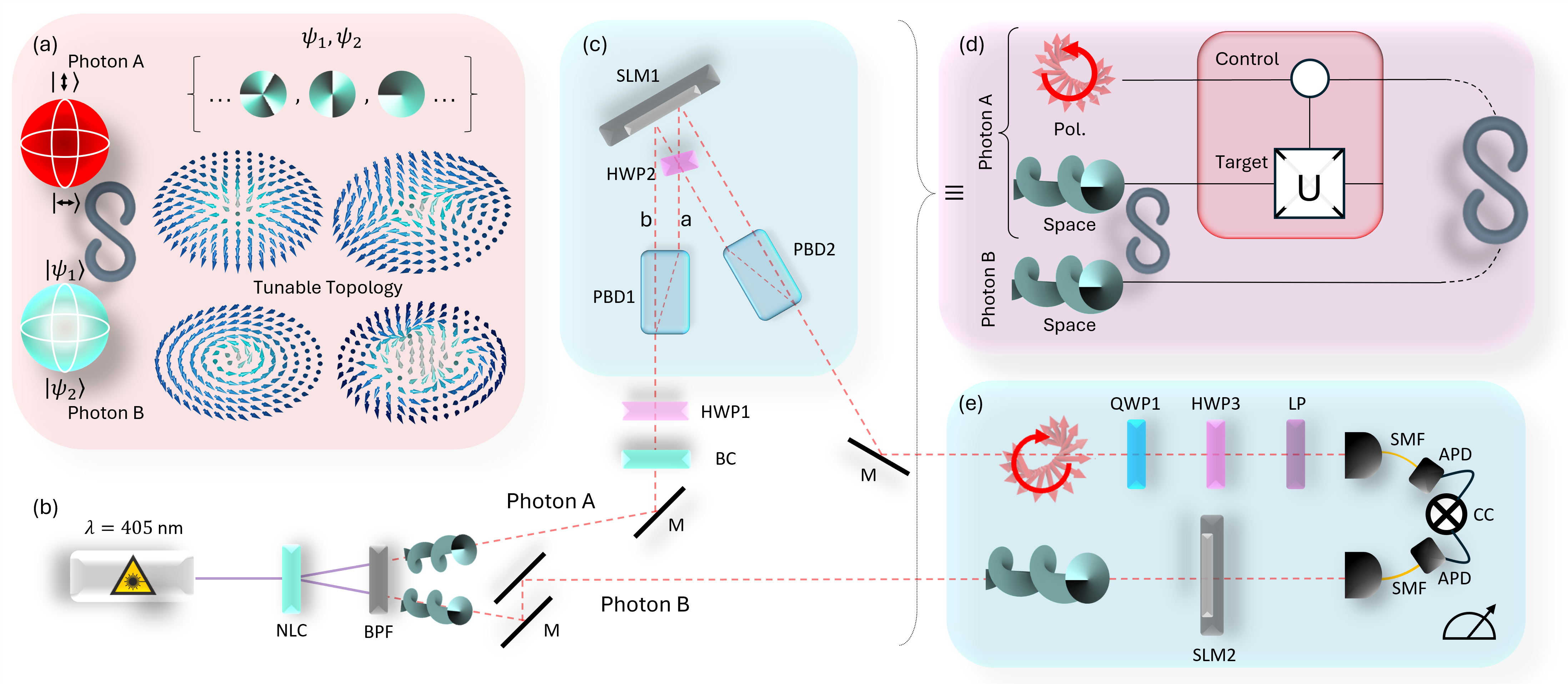}
	\caption{(a) Hybrid biphoton states possessing nonlocal correlations between the polarisation of one photon (Photon A) and the spatial mode of the other photon (Photon B), exhibit tunable topological structures.
    (b) Schematic for generating hybrid entanglement using SPDC states initially entangled in the OAM Dof. NLC: nonlinear crystal; BPF: band-pass filter; M: mirror; SLM: spatial light modulator; BC: BBO birefringent crystal; APD: avalanche photo diodes; HWP: half-wave plate; QWP: quarter-wave plate; LP: linear polarizer; PBD: polarization beam displacers; SMF: single-mode fiber; CC: coincidence counter.  (c) Digital self-locking Mach–Zehnder interferometer for performing the spatial-to-polarisation mode mapping constructed using two PBDs a HWP and a an SLM. A BBO crystal was used to compensate for the mode-dependent delay. (d) Circuit representation of the interferometer acting as a controlled unitary operator in the polarisation and spatial mode of photon A, performing a conditional unitary operation (OAM modulation), depending on the input polarisation state of photon A. (e) At the end of this process, photon A is measured in the polarisation DoF using a QWP, HWP and LP  while photon B is measured in the spatial DoF using SLM2.}
	\label{fig:schematic}
\end{figure*}

\noindent A schematic of the setup for generating hybrid entangled states that carry tunable skyrmionic topologies is shown in Fig.~\ref{fig:schematic}. Here, the skyrmionic topology emerges as a shared property of our two entangled photons, illustated as spin textures in Fig. \ref{fig:schematic}(a), extracted from the mapping between the real space of photon B ($R^{2}$) and the polarisation (spin) state space of photon A ($S^{2}$) \cite{ornelas2024non}. These topologies are then characterized by the skyrmion number, quantifying the integer number of times the polarisation state space of photon A is wrapped when traversing all possible positions for photon B. Next, we outline the procedure for creating these states by mapping spatially entangled photons onto multiple hybrid entangled states with embedded topology using our reconfigurable circuit.

\noindent We present the initial stage of the process in Fig. \ref{fig:schematic} (b) where a UV laser operating at a wavelength of $405 \,\text{nm}$ impinges on a type-I PPKTP nonlinear crystal (NLC) producing photon pairs having identical polarisations and wavelengths ($\lambda = 810 \,\text{nm}$). The state describing the photon pairs can be expressed in the transverse spatial degree of freedom using the OAM basis,
\begin{equation}
    \ket{\psi}_{AB} = \sum_{\ell=-L}^{L} c_{|\ell|}  \ket{\ell}_A \ket{-\ell}_B \ket{H}_{A}\ket{H}_{B},
    \label{eq:OAMEnt}
\end{equation}
where $c_{|\ell|}$ are the coefficients for the bi-photon states $\ket{\ell}_A\ket{-\ell}_B$ with $\ket{\pm\ell}$ corresponding to transverse spatial basis modes that have a characteristic azimuth-dependent ($\phi$) wavefunction, $ \propto \exp( \pm i\ell \phi)$, with an associated OAM of $\pm \ell \hbar$ per photon. Here $\ket{H}_{A,B}$ labels the horizontal polarisation state of each photon, i.e. photon A and B, respectively. At this stage, the entangled state occupies the $D = (2L+ 1)^2$  dimensional space,  $\mathcal{H}^{AB} 
= \mathcal{H}^{space}_A \otimes \mathcal{H}^{space}_B$. Next, photon A is transmitted through the interferometer so the state becomes (ignoring the polarisation of photon B)
\begin{equation}
    \ket{\psi}_{AB} \rightarrow \sum_{\ell=-L}^{L} c_{|\ell|} \left(  \ket{V, \ell+\ell_1}_A + \ket{H, \ell+\ell_2}_A  \right) \ket{-\ell}_B,
    \label{eq:OAMEntmapped}
\end{equation}
 for $\ell_1 \neq \ell_2$, resulting from the mapping 
\begin{align}
\ket{V, \ell} &\rightarrow \ket{H, \ell+\ell_{1}}, 
\nonumber \\ 
\ket{H, \ell} &\rightarrow \ket{V, \ell+\ell_{2}},
\label{eq:mapping}
\end{align}
 represented in the linear polarisation basis with  $\ket{V}$ corresponding to the vertical polarisation state.  
Coupling of photon A into a single-mode fiber (SMF) collapses the OAM state of photon A onto the state $\ket{\ell=0}$, thereby setting the condition, $\ell+\ell_1 =0$ or $\ell+\ell_2 = 0$. Therefore, correlations between the two photons are only observed when photon B is projected onto the OAM states $\ket{-\ell_1}$, $\ket{-\ell_2}$ or superpositions thereof.
 As a result, the post-selected state becomes the desired hybrid entangled state
\begin{equation}
    \ket{\Phi}_{AB} =\frac{1}{\sqrt{2}}\left( \ket{H}_A \ket{\ell_1}_B + \ket{V}_A\ket{\ell_2}_B \right),
    \label{eq:target}
\end{equation}
occupying the  $\mathcal{H}^{AB}_4 = \mathcal{H}^{pol}_2 \otimes \mathcal{H}^{space}_2$ Hilbert space where  photon A occupies the subspace $ \mathcal{H}^{pol}_2 = \text{span}\{\ket{H}, \ket{V}\}_{pol}$ spanned by polarisation states and photon B occupies the subspace $\mathcal{H}^{space}_2 = \text{span}\{\ket{\ell_1}, \ket{\ell_2}\}_\text{spatial}$ spanned by OAM states. Photon A is restricted to a two-dimensional subspace, however, multiple subspaces can be formed for photon B by simply selecting $\ell_1$ and $\ell_{2}$ since they can be any integer, i.e. $\ell_{1,2} \in \mathbb{Z}$. Consequently, we can generate various quantum skyrmion topologies on demand via the control of these subspaces, giving us access to many topological classes. Embedded in Eq.~(\ref{eq:target}), are the quantum topologies, a shared feature of the two photons, that can be characterised by the topological invariant (skyrmion number)
\begin{equation}
N =  p_{12} \Delta \ell,
\label{eq:SkyrmeNo2}
\end{equation}
where $p_{12} = \text{sgn}\left(|\ell_{1}| - |\ell_{2}|\right)$ and $\Delta \ell_{12} = \ell_1 - \ell_2$ are the polarity and the vorticity of the joint state, describing global features of their joint spatial-polarization correlations. 
The polarity of the state indicates the behaviour of the transition between polarization basis states $\{\ket{H}_A, \ket{V}_A\}$ with changing radial position measured for photon B. For example, $p_{12} = 1$ if photon A is in state $\ket{V}_A$ when photon B is measured at $r=0$ and in state $\ket{V}_A$ when photon B is measured at $r\to\infty$ and $p_{12} = -1$ if the opposite holds. Furthermore, the vorticity reveals attributes of the correlations when different angular positions for photon B are measured. The magnitude of the vorticity indicates the number of times the polarization state of photon A circles the poles of its state space with one complete cycle of photon B's azimuthal position, whereas the sign of the vorticity indicates the direction in which photon A's state rotates with increasing azimuthal position, $+1$ anti-clockwise and $-1$ clockwise. Therefore by tuning $\ell_{12}$ we can select between various topological classes and textures within each class, offering a general avenue to tune between different quantum biphoton state topologies.

To achieve the desired mappings that produce our quantum topologies, we engineer spatial-to-polarisation conversion optics as presented in Fig.~\ref{fig:schematic} (c) showing the self-locking Mach-Zehnder interferometer, which includes two beam displacers and a spatial light modulator (SLM) digitally split into two holograms that are controlled independently \cite{wu2021heralded}. This system can enact the transformation rules depicted in Eq. (\ref{eq:mapping}). 

Next, we convince the reader that the interferometer can indeed perform the desired transforms on photon A, specifically, the key transforms that enable photon A to eventually occupy the state in Eq.~(\ref{eq:OAMEnt}) to Eq.~(\ref{eq:OAMEntmapped}). To begin, we assume the polarisation state of the photon is first mapped to the diagonal polarisation state with a half-wave plate, 
\begin{equation}
\ket{H, \ell}_A \xrightarrow{\text{HWP}_1} \ket{V, \ell} _A+ \ket{H, \ell}_A.
\end{equation}
Subsequently, a polarising beam displacer ($\text{PBD}_1$) separates the $H$ and $V$ components into paths $a$ and $b$, followed by a rotation of the polarisation on path $a$ using the half-waveplate, $\text{HWP}_2$:
\begin{equation}
\ket{V, \ell}_A+ \ket{H, \ell}_A \xrightarrow{\text{PBD}_1+\text{HWP}_2} \ket{H, \ell, a}_A + \ket{H, \ell, b}_A.
\end{equation}
On the spatial light modulators (SLMs), OAM shifts of $\ell_{1}$ and $\ell_{2}$ are applied to paths $a$ and $b$, respectively. Path $b$ is then converted back to the $V$ polarisation (using $\text{HWP}_2$) and the two paths are recombined with a second PBD ($\text{PBD}_2$) so that the two path modes are collapsed onto a single path after $\text{PBD}_2$.  The steps are summarised as
\begin{align}
&\xrightarrow{\text{SLM}} \ket{H, \ell+\ell_{1}, a}_A + \ket{H, \ell+\ell_{2}, b}_A, \nonumber \\
&\xrightarrow{\text{HWP}_2 + \text{PBD}_2} 
  \ket{H, \ell+\ell_{1}}_A + e^{i \chi}\ket{V, \ell+\ell_{2}}_A,
\end{align}
where the phase $\chi$, arises from path mismatch and can be compensated digitally on the SLM.  This confirms the key transformation of photon A from its initial state in Eq.~(\ref{eq:OAMEnt})  to the state in Eq.~(\ref{eq:mapping}) before the filtering step with the single-mode fiber.
 \begin{figure*}[t]
\centering\includegraphics[width=1\linewidth]{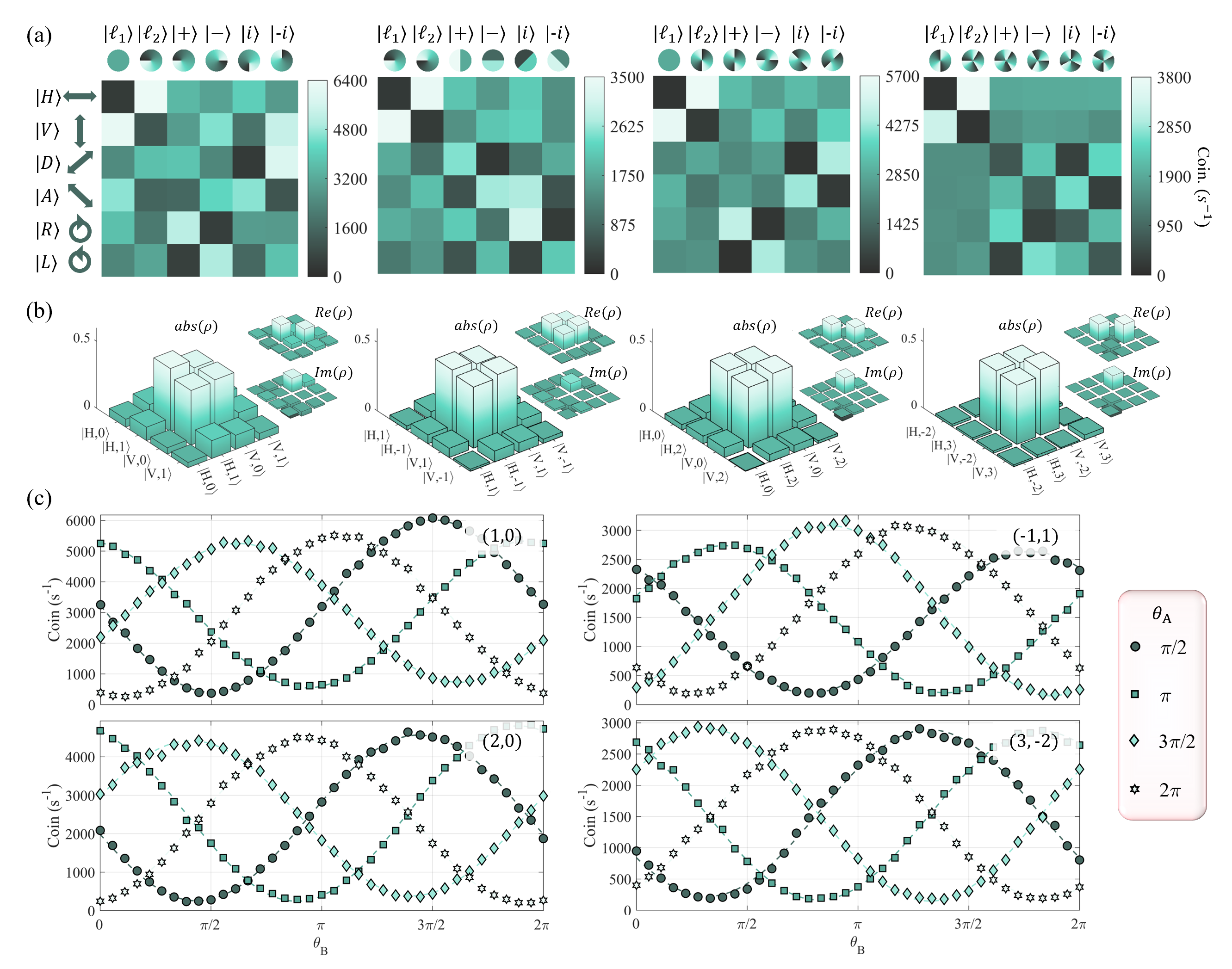}
	\caption{State reconstruction and characterisation across multiple subspaces $(\ell_{1}, \ell_{2})$ for states of the form$ \ket{\Phi}_{AB} = \ket{H}\ket{\ell_1} + \ket{V}\ket{\ell_2}$. (a) Joint projective measurement outcomes on photon A (polarization, rows) and B (OAM, columns) with their corresponding (b) reconstructed density matrices for the states with $(\ell_1,\ell_2) \in \{(1,0), (-1,1), (2,0), (3,-2)\}$. The magnitude $(\text{abs}(\rho))$ $(\text{Re}(\rho))$ and imaginary, $(\text{Im}(\rho))$, components of the density matrix are shown. (c) Joint detection measurement outcomes for the  CHSH Bell-inequality test for subspaces $(1,0)$, $(1,-1)$, $(2,0)$, and $(3,-2)$ respectively, with the associated angles for photon A ($\theta_{A}$) shown in the legend below. The measured subspaces admit visibilities of $V = 0.86, 0.90, 0.91, 0.91$, and Bell parameters, $S = 2.35\pm0.02, 2.49\pm0.02, 2.56\pm0.02 , 2.56 \pm0.03$,  respectively.}
	\label{fig:tomo}
\end{figure*}
Lastly, this whole transformation acts as a conditional spatial-mode shaper, analogous to a controlled-unitary gate (up-to a unitary rotation in the polarisation basis) where the photon’s independent internal degrees of freedom serve as control and target subsystems, as depicted in Fig.~\ref{fig:schematic} (d). Based on the transformation rules for the input and output states in Eq. (\ref{eq:mapping}), we can express the underlying operator as 
\begin{align}
    \mathcal{T} &=  |H\rangle\langle V| \otimes \ket{\ell+\ell_1}\bra{\ell} + |V\rangle\langle H| \otimes \ket{\ell+\ell_2}\bra{\ell},  \nonumber\\
     &= \sigma_x |H\rangle\langle H| \otimes \mathcal{U}_{\ell_2} + \sigma_x |V\rangle\langle V| \otimes \mathcal{U}_{\ell_1},
\end{align}
showing that the operator applies the unitary operators $\mathcal{U}_{\ell_1} = \ket{\ell+\ell_1}\bra{\ell}$ or $\mathcal{U}_{\ell_2} = \ket{\ell+\ell_2}\bra{\ell}$ conditioned on  the input polarisation, $V$ or $H$, respectively. Here $\sigma_x$ is the bit flip operator in the polarisation basis. Importantly, this action is reprogrammable since the values for $\ell_{1,2}$ can be controlled digitally and independently. Furthermore, we emphasize that whilst the OAM basis is a convenient basis within which to express our entangled states, spatial light modulators are capable of performing any spatial transformation on our photons, thus suggesting our scheme is able to perform a general controlled-unitary transformation.

\section*{Results}
\begin{table}[h!]
\centering
\caption{Summary of quantitative measurements for different subspaces $(\ell_1, \ell_2)$; Fidelity ($F$),  Purity ($\gamma$), Concurrence ($C$)) and Skyrmion number ($N$).}
\begin{tabular}{|c|c|c|c|c|}
\hline
$(\ell_1, \ell_2)$ & F & $\gamma$ & $C$ & $N$ \\
\hline
(1,0) & 0.8527  & 0.8213  & 0.8414 & 0.9932\\
\hline
(-1,0) & 0.8622  & 0.8231  & 0.8095 & -0.9936\\
\hline
(2,0) & 0.9296  & 0.8889  & 0.9101 & 1.9987\\
\hline
(-2,0) & 0.8569  & 0.8678  & 0.8299 & -1.9987\\
\hline
(3,0) & 0.9006  & 0.8530  & 0.8519 & 2.9971\\
\hline
(-3,0) & 0.8812  & 0.8515  & 0.8439 & -2.9972\\
\hline
(1,-1) & 0.8564  & 0.8850  & 0.8560 & 0.0027\\
\hline
(2,1) & 0.8512  & 0.8262  & 0.8055 & 0.9880\\
\hline
(-3,1) & 0.8019  & 0.8513  & 0.8496 & -3.9957\\
\hline
(3,-1) & 0.8357  & 0.8528  & 0.8469 & 3.9957\\
\hline
(-3,2) & 0.8338  & 0.8399  & 0.8328 & -4.9037\\
\hline
(3,-2) & 0.9338  & 0.8764  & 0.8780 & 4.9037\\
\hline
\end{tabular}
\label{table:quantifiers}
\end{table}
We experimentally generated and measured several hybrid states with different post-selected subspaces, $(\ell_1, \ell_2)$, of various symmetries (anti-symmetric, $\ell_1 = -\ell_2$) or (asymmetric, $|\ell_1| \neq |\ell_2|$) for variability.  The different hybrid states were post-selected by encoding different phase holograms on the two halves of SLM 1. To improve the purity of the generated states, a blazed grating was added to the holograms. To measure the states, we first performed state tomography using the hybrid degrees of freedom to verify that we can reliably create the target states in Eq. (\ref{eq:target}) for several values of $\ell_1$ and $\ell_2$. Some measurement outcomes are shown in Fig.~\ref{fig:tomo} (a) for the states with $(\ell_1,\ell_2)\in\{(1,0),(-1,1),(2,0),(3,-2)\}$. We can represent the overcomplete projection measurements that were carried out sequentially on photon A and B using the computational basis as, $\{\ket{0}, \ket{1}, \ket{+}, \ket{+i}, \ket{-}, \ket{-i}\}$, where the unnormalised superposition states are $\ket{\pm} = \ket{0} \pm \ket{1}$ and $\ket{\pm i} = \ket{0} \pm i \ket{1}$ assuming that $\{ \ket{H} \equiv \ket{0},  \ket{V} \equiv \ket{1}  \}$ and $\{ \ket{\ell_1} \equiv \ket{0},  \ket{\ell_2} \equiv \ket{1} \}$ for photon A and B,  respectively. These projections probe the $d^2 =4$ dimensional Hilbert space,  $\mathcal{H}_4 = \mathcal{H}^{pol}_2 \otimes \mathcal{H}^{space}_2$,  of photon A and B.  The measurement outcomes were thereafter used to reconstruct the state by employing a maximum likelihood algorithm, \cite{jack2009precise} that minimises the square error between the experimental joint measurements and the modelled measurements, assuming that the underlying state has the decomposition \cite{agnew2011tomography} 
$\rho = \sum^{15}_{m,n=0} b_{m} \Gamma_{m}$, where  $\Gamma_{m}$ represents the generalised Gellman matrices in 4 dimensions including the identity $\Gamma_0 \equiv \mathbb{I}_4$.  The reconstructed density matrices are shown in Fig.~\ref{fig:tomo} (b).  To quantify the similarity between the experimental and ideal states, we measured the \textit{state fidelity}, which is a metric that scales from 0 (not matching) to 1 (identical) and is defined as
$F(\rho, \sigma) = \left( \mathrm{Tr}\left[\sqrt{\sqrt{\rho} \, \sigma \, \sqrt{\rho}} \right] \right)^2$. A summary of the fidelities for all the measured states can be seen in Table \ref{table:quantifiers}, showing measured fidelities  $0.8 < F(\rho, \sigma) \approx0.93$.  The fidelities can be further improved by performing aberration correction on the SLMs in the interferometer to reduce crosstalk. Furthermore, we note the presence of a consistent relative phase of $\frac{\pi}{6}$ in every measurement. Whilst this can partially be attributed to the path mismatch, we also note that the use of dielectric mirrors within the experimental setup can also result in a change in the polarization basis, introducing some relative phase between the horizontal/vertical basis states. Beyond this constant accumulated phase, we also identify a mode dependent phase that can be attributed to Gouy phase, accumulated in the experiment. This resulted in the approximate phases, $\{-\frac{3\pi}{8},-\frac{\pi}{6},-\frac{14\pi}{24},-\frac{3\pi}{8}\}$,  identified for the subspaces, $(\ell_1, \ell_2) = (1,0), (1,-1), (2,0), (3,-2)$, respectively. 
Next, we quantified the purity and degree of coherence, $\gamma = \text{Tr}(\rho^2)$, which is also related to the linear entropy ($S_L =1-\gamma$) of the states. State purity scales from $\gamma = 0.25$ (two qubits) for maximally mixed states to $\gamma = 1$ for pure states, therefore, states of high purity indicate a high degree of coherence in the given subspace. The measured purities for our states are summarised in Table \ref{table:quantifiers}, showing measurements ranging between  $ 0.82\leq \gamma \leq 0.89$. Because the purity of a state is associated with its coherence, these values can be improved by reducing the background noise in the experiment, reducing the decoherence effects introduced by the polarisation-dependent mode delays in our interferometer as-well as reducing the gating time in the experiment (a $ 2 \ ns$ gating time was used). The spectral purity of the photons produced via SPDC can also contribute to the overall purity of the state (a $10 \text{nm}$ wavelength bandpass filter, centred on $810 \text{nm}$ was used).

\begin{figure*}[t]
\centering\includegraphics[width=1\linewidth]{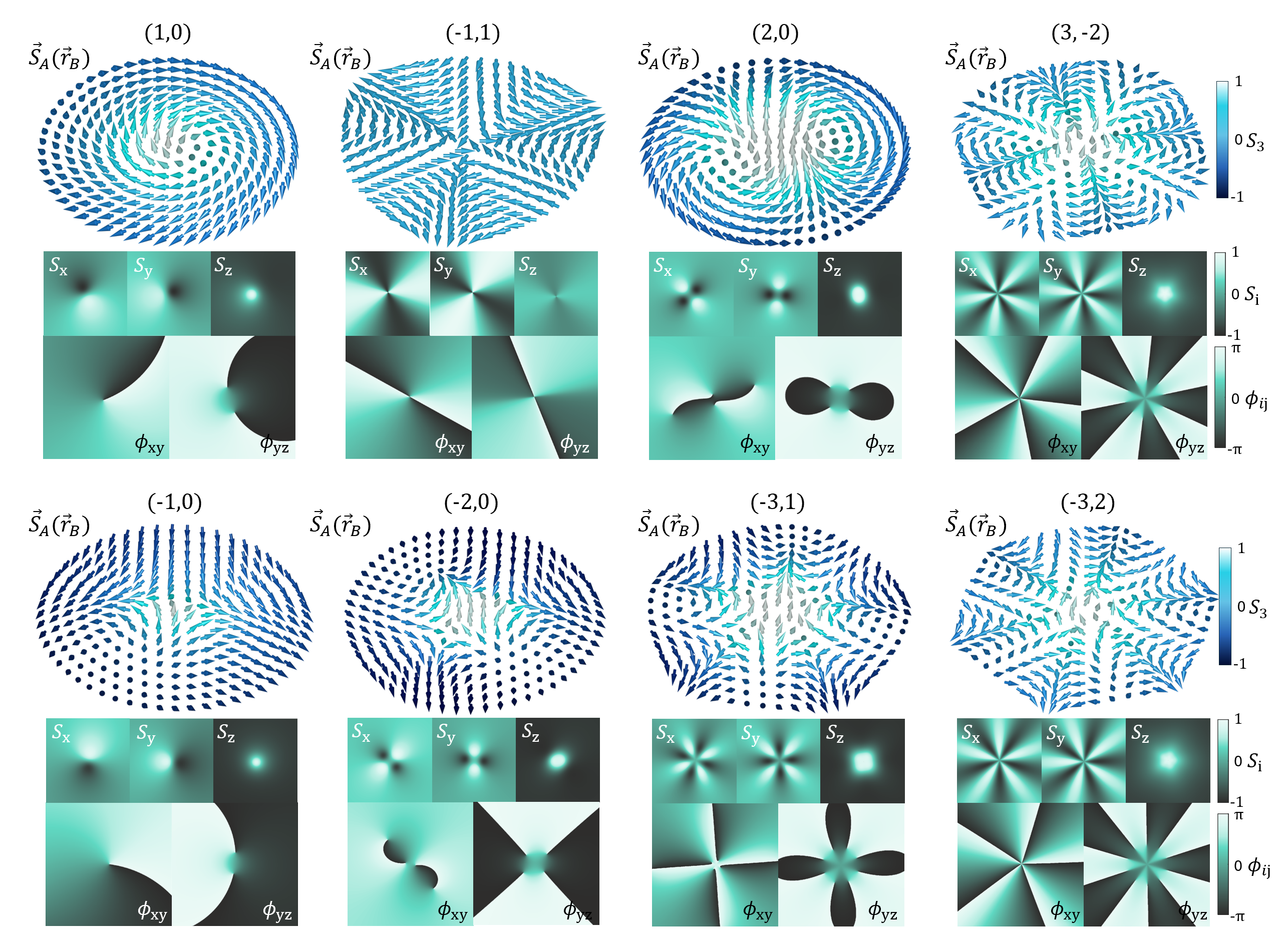}
	\caption{The reconstructed quantum Stokes vector field, $\vec{S}_A(\vec{r}_B)$, for every state with the associated vector components, $S_{x,y,z}$ and Stokes phases $\phi_{ij}=\tan^{-1}\left(\frac{S_j}{S_i}\right)$, for states with topological numbers $N\in \{0.9932, 0.0027, 2.9971, 4.90\}$ found in the upper panel and $N\in \{-0.9932, -1.9987, -3.9957, -4.9037\}$ found in the lower panel. 
    } 
	\label{fig:Top}
\end{figure*}

Next, we proceeded to characterise the quantum correlations by measuring the degree of entanglement/nonseparability, namely concurrence $C$ \cite{wootters2001entanglement}, of our reconstructed density matrices and report them in Table \ref{table:quantifiers}. The concurrence was measured using $ C(\rho) = \text{max}\{0,\lambda_1-\sum_{j=2}^4\lambda_j\},	$ where
$\lambda_j$ are the eigenvalues of the operator $R = \sqrt{\sqrt{\rho} \tilde{\rho} \sqrt{ \rho }}$ in descending order of magnitude and $\tilde{\rho}= \sigma_y\otimes \sigma_y \rho^* \sigma_y \otimes \sigma_y$.   Here  $^*$ denotes a complex conjugation, while $\sigma_y$ is the Pauli-y gate.  For reference, separable (pure or maximally mixed) states have a concurrence  $C = 0$ , while nonseparable and entangled states have concurrence  $C = 1$. We emphasize here that typical SPDC experiments produce OAM entangled states with probabilities which vary depending on mode order. This automatically reduces the achievable degree of non-separability for asymmetric OAM states, necessitated for topologically structured states. Ordinarily, this may be circumvented through pump shaping or procrustean filtering at the cost of complicating or introducing further loses to the experiment. Instead, by tuning the angle of $\text{HWP}_1$ we compensate for the mismatch in probabilities between asymmetric states without introducing any additional complexity or signal loss. Practically, the angle of $\text{HWP}_1$ was chosen so as to equalize the coincidences measured for each basis state projection, i.e, to equalize the coincidence counts found within the first $2\!\times\!2$ measurement block in the QST measurement.. For our system, the measured concurrences are within the range  $0.81\leq C \leq 0.88$ for the tested subspaces, showing the states are indeed entangled and nonseparable with our current configuration. 

A strict measure for nonclassical correlations, specifically nonlocality, is the violation of the Clauser–Horne–Shimony–Holt (CHSH) Bell inequality \cite{clauser1969proposed}. To verify non-classicality, we performed correlation measurements using joint superposition projections on states, $\ket{\theta_A} \propto \ket{H}+ \exp(i\theta_A) \ket{V}$ and $\ket{\theta_B} \propto \ket{\ell_1}+ \exp(i\theta_B) \ket{\ell_2}$, where the relative phases, $\theta_A$ and $\theta_B$ determine the states. For a maximally entangled state, the coincidence counts are proportional to $ J(\theta_A, \theta_B) = |\langle \theta_A | \langle \theta_B | \Psi  \rangle|^2  \propto \cos^2 \left(\frac{(\theta_A-\theta_B)}{2}  \right)$, consistent with our experimental coincidence measurements as shown in Fig.~\ref{fig:tomo} (c). We quantified the average visibilities for each case, ($V = ( J_{max} - J_{min}) / ( J_{max} + J_{min})$) and found that they are within the range, $ 0.86\leq V \leq 0.91$. We note that to obtain a high degree of visibility a birefringent crystal (BC) was used to compensate for any optical path delay (OPD) introduced by our interferometer. The  coherence length of the entangled photons is $L_C = 66 \,\mu m$ (obtained approximately using $L_C\approx\frac{\lambda^2}{\Delta \lambda}$) which means that minor OPD contributions from the blazed grating on the SLM, the tilt angle of HWP2 and minor manufacturing inconsistencies between the two PBDs may contribute significantly to a drop in the measured state's visibility.


Furthermore, we measured the Bell-parameter $(S_B)$ from the coincidence counts, without any accidental subtraction, for the states shown in Fig.~\ref{fig:tomo}. We found that they are within the range $2.35 \leq S_B \leq 2.56$  for all  states tested. The measured values are all above the classical limit of $S = 2$ for local hidden variable theories, indicating that the measured topological hybrid states are genuinely nonlocal. While we manage to violate the Bell parameter for every state shown in Fig.~\ref{fig:tomo} (c), an eventual decrease in the measured Bell parameter is observed when the OAM mode orders are increased. This can be attributed to the reduction in counts observed for higher-order modes. This reduction arises from two factors: the coupling efficiency decreases with increasing OAM charge ($\ell_{12}$), and the spontaneous parametric down-conversion (SPDC) process itself produces fewer photons in modes of higher topological charge \cite{miatto2011full}. Together, these effects lower the signal-to-noise ratio, manifesting as a decay in the measured visibilities, with the lowest value of $V \approx 0.80$ observed for $\ell_{12} = \pm 4$.\\

\noindent Finally, the topology of each state was characterized by calculating the quantum Skyrmion number, 
\begin{equation}
N=\frac{1}{4\pi}\int\int_{R^2} \vec{S}_A(\vec{r}_B) \cdot \left( \partial_x\vec{S}_A(\vec{r}_B) \times \partial_y\vec{S}_A(\vec{r}_B) \right) \, d\vec{r}
\end{equation}
where $\vec{S}_A(\vec{r}_B)$ is the non-local quantum Stokes vector field given by, $\vec{S}_A(\vec{r}_B) = \text{Tr}\left(\vec{\sigma}\,\langle \vec{r}_B|\rho | \vec{r}_B\rangle \right)$ with $\vec{\sigma}=\begin{bmatrix}\sigma_x & \sigma_y & \sigma_z \end{bmatrix}^T$ and $\sigma_i$ being the usual Pauli-spin matrices. The full set of measured Skyrmion numbers are summarised in Table \ref{table:quantifiers}, showing measurements matching closely to the expected integer values. A collage of various topological structures for different measured states with diverse topologies are shown in Fig.~\ref{fig:Top}. Here, the vector field $\vec{S}_A(\vec{r}_B)$ along with the vector components, $S_{x,y,z}$ and Stokes phases $\phi_{ij}=\tan^{-1}\left(\frac{S_j}{S_i}\right)$ depict different aspects of some of the diverse topological structures achievable within our modular system. By varying the vorticity parameter, $v=\ell_1-\ell_2$, we achieved states belonging to 11 distinct topological classes characterized by topological numbers ranging from $N=-5$ to $N=5$. This change in vorticity can alter both the sign and magnitude of the topology of the state. The variation in the sign and magnitude of the vorticity is visually captured by the change in the rotation direction of the texture about the $S_z$ axis ($+1$ if vectors rotate in the same direction as the changing azimuthal angle and $-1$ otherwise) and the change in the number of rotations performed by $\vec{S}_A(\vec{r}_B)$ about the $S_z$ axis, respectively. Furthermore, we note that the charge of the accompanying Stokes phases, in particular $\phi_{xy}$, can also be used to uniquely identify the change in vorticity with the Stokes parameter, $S_z$, indicating the polarity ($+1$ for $\vec{S}_A(\vec{r}_B)$ pointing up in the centre and down away from the origin and $-1$ for the opposite scenario). We note further that the accompanying polarization singularities identified in the Stokes phases also highlight the ability of our system to generate hybrid states with diverse and tunable 1D topological structures, distinct from the Skyrmionic structures emphasized in this work. Whilst we have shown states with tunable topology through altering their vorticity, we can also tune their polarity by simply switching the holograms shown in $\text{SLM}_1$ and further by introducing a pair of waveplates after the interferometer we could also alter the polarization basis of the states thereby switching between bimeronic \cite{shen2021topological} and Skyrmionic topological structures.


 
 \section{Discussion and Conclusion}
In conclusion, we presented a method for structuring quantum states of light to produce genuine nonlocal spatial-polarisation hybrid entangled states that show high violations of the Bell inequality. This was achieved using a self-locking Mach–Zehnder interferometer that modulates one photon from a pair of photons produced via SPDC. We showed the interferometer implements the required mode mappings, effectively enacting a controlled unitary operation between the polarisation and spatial degrees of freedom of a single photon. By incorporating a spatial light modulator within the interferometer, the system becomes digitally reprogrammable. Additionally, we note that whilst we have demonstrated the ability of our system to generate arbitrary topological states from initial OAM-OAM entangled state, our system can also be used to perform the same transformation on initially polarization entangled states thereby enhancing the versatility of our scheme. 
This advances methods for generating high-quality, structured quantum light in a controlled manner, suitable for creating numerous entangled topological states of light. Recent advances in reprogrammable logical gates have gained prominence by exploiting the combined control of multiple photonic degrees of freedom, such as polarization and OAM \cite{wang2024polarization, ru2021realization, marques2026experimental}. Our work contributes directly to this growing body of methods by demonstrating a flexible approach for hybrid-state generation within the same framework. Importantly, since hybrid entanglement synthesis is a crucial ingredient for the creation of quantum skyrmions \cite{ornelas2024non}, i.e., topological states known for their robustness to noise \cite{ornelas2025topological}, our method provides a powerful pathway for producing such states with high quality. In this way, the reprogrammable gate paradigm not only extends current strategies for structured photonic entanglement but also opens new opportunities for exploring topologically protected quantum information processing, using digital approaches.

\section*{Acknowledgements} 
AF and IN acknowledge financial support from the DSI-CSIR Rental Pool Programme and  the NRF. PO acknowledges financial support from the DSTI-CSIR Inter-programme Bursary Scheme. AGO acknowledges financial support from the Coordena\c{c}\~ao de Aperfei\c{c}oamento de Pessoal de N\'ivel Superior—Brasil (CAPES), Finance Code 001. CRG acknowledges financial support from Secihti through the project CBF-2025-I-1804.

\section*{References}

\providecommand{\newblock}{}

\end{document}